\documentclass[12pt,epsf]{article}
\pdfoutput=1
\usepackage[dvips]{graphicx}
\usepackage{verbatim,graphics,graphicx,color,slashed}
\usepackage{bm}

\begin{document}
\thispagestyle{empty}
\vspace*{-15mm}

\vspace{15mm}
\begin{center}
{\Large\bf
LHC Searches for Top-philic Kaluza-Klein Graviton\\
}
\vspace{7mm}

\baselineskip 18pt
{\bf Chao-Qiang Geng${}^{1, 2, 3*}$, Da Huang${}^{4\dagger}$, Kimiko Yamashita${}^{2, 3\ddagger}$}
\vspace{2mm}

{\it
${}^{1}$Chongqing University of Posts \& Telecommunications, Chongqing, 400065, China\\
${}^{2}$Physics Division, National Center for Theoretical Sciences,\\ Hsinchu, Taiwan 300\\
${}^{3}$Department of Physics, National Tsing Hua University, Hsinchu, Taiwan 300\\
${}^{4}$Institute of Theoretical Physics, Faculty of Physics, University of Warsaw, Pasteura 5, 02-093 Warsaw, Poland
\newline \newline
${}^{*}$geng@phys.nthu.edu.tw,
${}^{\dagger}$dahuang@fuw.edu.pl,
${}^{\ddagger}$kimikoy@phys.nthu.edu.tw}\\
\vspace{10mm}
\end{center}
\begin{center}
\begin{minipage}{14cm}
\baselineskip 16pt
\noindent
\begin{abstract}
We study the phenomenology of a massive graviton $G$ with non-universal couplings to the Standard Model (SM) particles. Such a particle can arise as a warped Kaluza-Klein graviton from a framework of the Randall-Sundrum extra-dimension model. In particular, we consider a case in which $G$ is top-philic, i.e., $G$ interacts strongly with the right-handed top quark, resulting in the large top-loop contributions to its production via the gluon fusion and its decays to the SM gauge bosons. We take into account the constraints from the current 13~TeV LHC data on the channels of $t\bar{t}$, $\gamma\gamma$, $jj (gg)$, $\gamma Z$, and $ZZ$.
Consequently, it is found that the strongest limit for this spin-2 resonance $G$ comes from the $t\bar{t}$ pair search, 
which constrains the cutoff scale to be of ${\cal O}$(100~GeV) for the right-top coupling of ${\cal O}(1)$
and the massive graviton mass in the range $m_G=2$--$5$~TeV, 
significantly relaxed compared with the universal $G$ coupling case. 
\end{abstract}
\end{minipage}
\end{center}

\baselineskip 18pt
\def\thefootnote{\fnsymbol{footnote}}
\setcounter{footnote}{0}

\newpage

\section{Introduction}
Randall-Sundram (RS) model~\cite{Randall:1999ee} has an attractive feature as
providing an explanation of the gauge hierarchy, $\mathcal{O}(10^{16})$, between the reduced Planck mass  and  electroweak scales.
Not only solving the puzzle, it also has a nice potential  to connect dark matter~\cite{Lee:2013bua, Lee:2014caa, Kraml:2017atm, Rueter:2017nbk}
as its prediction of the Kaluza-Klein (KK) graviton naturally interacts with all of particles via each energy-momentum tensor.

It is known that the RS model of the universal coupling case (i.e. the KK graviton interacts with particles with the same coupling strength)
suffers from strong constraints on the model parameters from the current LHC experimental results~\cite{Kraml:2017atm}.
It is necessary to reconsider the non-universal case of the RS model (\cite{Goldberger:1999wh,Davoudiasl:1999tf,Pomarol:1999ad,Chang:1999nh,Davoudiasl:2000wi,Dillon:2016fgw,Dillon:2016tqp,Dillon:2017ctw}) 
from the latest LHC experimental data and explore  the corresponding constraints  for the model.
One of such a case is that  the KK graviton interacts strongly with the top quark, in which
it was found that top-loop effects can be  comparable with the tree-level ones for the KK graviton productions and its decays~\cite{Geng:2016xin}.
For simplicity, we concentrate on that only the right-handed top quark interacts with the KK graviton
 via a coupling of ${\cal O}(1)$, whereas
 the profile of the left-handed top quark is far away (UV brane) from a KK graviton wave function, which is localized near the IR brane.
We assume that the color $SU(3)_c$ and hypercharge $U(1)_Y$ gauge fields are in the bulk, so that 
the couplings with the KK gravitons are diluted by a volume factor.
The other Standard Model (SM) particles are localized near the UV brane.
For this setup, our signals are $t\bar{t}$, $\gamma\gamma$, $jj (gg)$, $\gamma Z$, and $ZZ$.
We will derive the constraints on the model parameters 
from  the latest 13~TeV LHC searches for the KK graviton resonance decaying to these final states.

The article is organized as follows.
Our model is presented in Sec.~\ref{sec:model}, and
the effective couplings  by top-loops are described in Sec.~\ref{sec:e_coupling}.
The KK graviton productions and decays are shown in Sec.~\ref{sec:production} and Sec.~\ref{sec:decay}, respectively.
Constraints on the model parameters by the current 13~TeV LHC are  discussed in Sec.~\ref{sec:limit}.
The summary is given in Sec.~\ref{sec:summary}.

\section{Top-philic KK Graviton Model}\label{sec:model}
Besides being a possible solution to the gauge hierarchy problem,  the generalized RS models provide us with a general framework to study the massive KK graviton with non-universal couplings to the SM fields. In this framework, the geometry is a slice of a five-dimensional (5D) warped spacetime with two boundaries corresponding to UV and IR branes, respectively. All of the SM fields are promoted to 5-dimensional objects, either propagating in the bulk or located on the branes.  The interactions among particles are given by the overlapping wave-functions of the involving particles, which naturally give rise to the hierarchy in the model couplings. In particular, the wave-function of the first KK graviton $G$ is peaked near the IR brane, so that the fields located on or around the IR brane would couple to this massive graviton strongly, while other fields near the UV brane would have exponentially smaller couplings with $G$. Concretely, we can write down the following general interactions between $G$ and SM particles
\begin{eqnarray}
{\cal L} = - \sum_i \frac{c_i}{\Lambda} G_{\mu\nu} T_i^{\mu\nu}\,,
\end{eqnarray}
where $T_i^{\mu\nu}$ denotes the energy-momentum tensor for the i-th SM particle  with $c_i$ the corresponding the coupling strength, 
and $\Lambda$ is the typical cutoff scale for the $G$ interactions. 
In the simple case when the 5D bulk geometry is the AdS$_5$ spacetime with its curvature $k$ and its length $L$, the mass and cutoff scale of this KK graviton is predicted to be $m_G\approx 3.8 k e^{-kL}$ and $\Lambda\sim \bar{M}_{\rm Pl} e^{-kL}$, respectively, where $\bar{M}_{\rm Pl}$ is the reduced Planck mass in the ordinary 4-dimensional spacetime.

In the present work, we consider a model in which only the right-handed top quark field sits  around the IR brane, and the gauge bosons $G_\mu$ and $B_\mu$ corresponding to the color $SU(3)_c$ and hypercharge $U(1)_Y$ gauge groups live in the bulk, while other SM fields, including the $SU(2)_L$ gauge and  SM Higgs doublet bosons, are placed close to or exactly on the UV brane. According to the naive dimensional analysis, 
$G$ is expected to interact with the right-handed top quark strongly, while weakly with $W^\pm$ and SM Higgs bosons and other fermions. Furthermore, the wave-functions of the zero-mode gauge fields are always predicted to be constant in the bulk, so that their couplings to $G$ would be suppressed by the volume factor of order  $1/\ln(\bar{M}_{\rm Pl}/M_{\rm IR}) \sim 1/(kL) \sim 0.03$ with the IR brane scale at ${\cal O}(\rm TeV)$. 
Note that this suppression factor has the similar order  as the one-loop ones of $\alpha_{(s)}/(4\pi)$  for the EW (color) gauge bosons, with $\alpha_{(s)}$ referring to the electromagnetic (strong) fine structure constant. In the light of this observation, the interacting Lagrangian relevant to the phenomenology of the KK graviton is given by
\begin{eqnarray}\label{LagB}
{\cal L}_G &=& -\frac{G_{\mu\nu}}{\Lambda} \Bigg[\frac{\alpha c_1}{4\pi} \left(\frac{1}{4}\eta^{\mu\nu} B^{\lambda\rho} B_{\lambda\rho} - B^{\mu\lambda}B^{\nu}_{\ \lambda} \right) + \frac{\alpha_s c_{gg}}{4\pi} \left(\frac{1}{4}\eta^{\mu\nu} G^{a\,\lambda\rho} G^a_{\lambda\rho} - G^{a\,\mu\lambda}G^{a\,\nu}_{\ \ \lambda} \right)\nonumber\\
&& + c_{tt} \left(\frac{i}{4}\bar{t}_R (\gamma^\mu D^\nu + \gamma^\nu D^\mu) t_R -\frac{i}{4} (D^\mu \bar{t}_R \gamma^\nu + D^\nu \bar{t}_R \gamma^\mu) t_R \right.\nonumber\\
&& \left.
- i\eta^{\mu\nu}[\bar{t}_R \gamma^\rho D_\rho t_R -\frac{1}{2}D^\rho(\bar{t}_R \gamma_\rho t_R)] \right) \Bigg]\,,
\end{eqnarray}    
where $\eta_{\mu\nu} = \mathrm{diag}(1, -1, -1, -1)$ is the Minkowski metric tensor, and
$D_\mu = \partial_\mu + i (2/3) g_1 B_\mu + i g_s G^a_\mu$ is the covariant derivative for the right-handed top quark field. 
 Note that in Eq.~(\ref{LagB}) we have explicitly picked up one-loop gauge factors in front of the corresponding massive graviton couplings to SM gauge bosons, in order to explicitly represent the aforementioned bulk volume suppression factors. With this convention, the couplings $c_1$ and $c_{gg}$ 
 are of ${\cal O}(1)$, resulting in  the coupling sizes of 
\begin{eqnarray}\label{PowerCounting}
c_{tt} > \frac{\alpha_s c_{gg}}{4\pi} \sim \frac{\alpha c_1}{4\pi} \gg c_{\rm others}\,.
\end{eqnarray}
After the spontaneous EW symmetry breaking, the original coupling between $G$ and the $U(1)_Y$ gauge field $B_\mu$ is divided to the couplings with the electromagnetic $A_\mu$ and weak $Z_\mu$ fields as follows
\begin{eqnarray}\label{LagA}
{\cal L}_{G} &\supset & -\frac{1}{\Lambda}G_{\mu\nu} \left[ \frac{\alpha c_{\gamma\gamma}}{4\pi} \left(\frac{1}{4}\eta^{\mu\nu} A^{\lambda\rho} A_{\lambda\rho} - A^{\mu\lambda}A^{\nu}_{\ \lambda} \right) + \frac{\alpha c_{Z\gamma}}{4\pi} \left(\frac{1}{4}\eta^{\mu\nu} A^{\lambda\rho} Z_{\lambda\rho} - A^{\mu\lambda}Z^{\nu}_{\ \lambda} \right) \right.\nonumber\\
&+& \left. \frac{\alpha c_{ZZ}}{4\pi} \left(\frac{1}{4}\eta^{\mu\nu} Z^{\lambda\rho} Z_{\lambda\rho} - Z^{\mu\lambda}Z^{\nu}_{\ \lambda} \right)\right]\,,
\end{eqnarray}  
where the couplings of $c_{\gamma\gamma, Z\gamma, ZZ}$ can be derived from that of  $c_1$ with the transformations
\begin{eqnarray}
c_{\gamma\gamma} &=& c_1 \cos^2 \theta_W\,,\quad c_{Z\gamma} = -c_1 \sin 2\theta_W = -\sin 2\theta_W c_{\gamma\gamma}/\cos^2\theta_W\,,\nonumber\\
 c_{ZZ} &=& c_1 \sin^2 \theta_W = \tan^2\theta_W c_{\gamma\gamma}\,,
\end{eqnarray}
which is a particular case studied in Ref.~\cite{Kim:2015vba}.
As a result, there are only three free parameters $c_{gg}$, $c_{tt}$ and $c_1$ to characterize the LHC signals of the KK graviton. 

Since the SM Higgs boson is placed far away from the IR brane, the model cannot solve the hierarchy problem with the large warped factor 
in the original RS proposals. 
Also we would like to point out that 
with the completely UV localized left-handed  and 
IR localized right-handed top quarks, it is difficult to generate the top quark mass~\cite{Ponton:2012bi,Csaki:2010az}. In fact, we do not require left-handed and right-handed top fields are strictly placed on the UV and IR branes, respectively, so that the overlap of these two fields in the bulk can generate top mass term, even though a large amount of tuning of 5D parameters is needed to achieve its observed large value.
However, our focus here is the LHC phenomenology for the top-philic KK graviton $G$ with the emphasis on the unconventional power counting rules 
 and the possible significance of the top-quark loop contributions to the productions and decays of $G$. The present model provides the minimal setup to realize this scenario.

\section{Top-loop Effects to Effective Couplings of $G$}\label{sec:e_coupling}
\begin{figure}[t!]\center
 \includegraphics[width=0.8\textwidth]{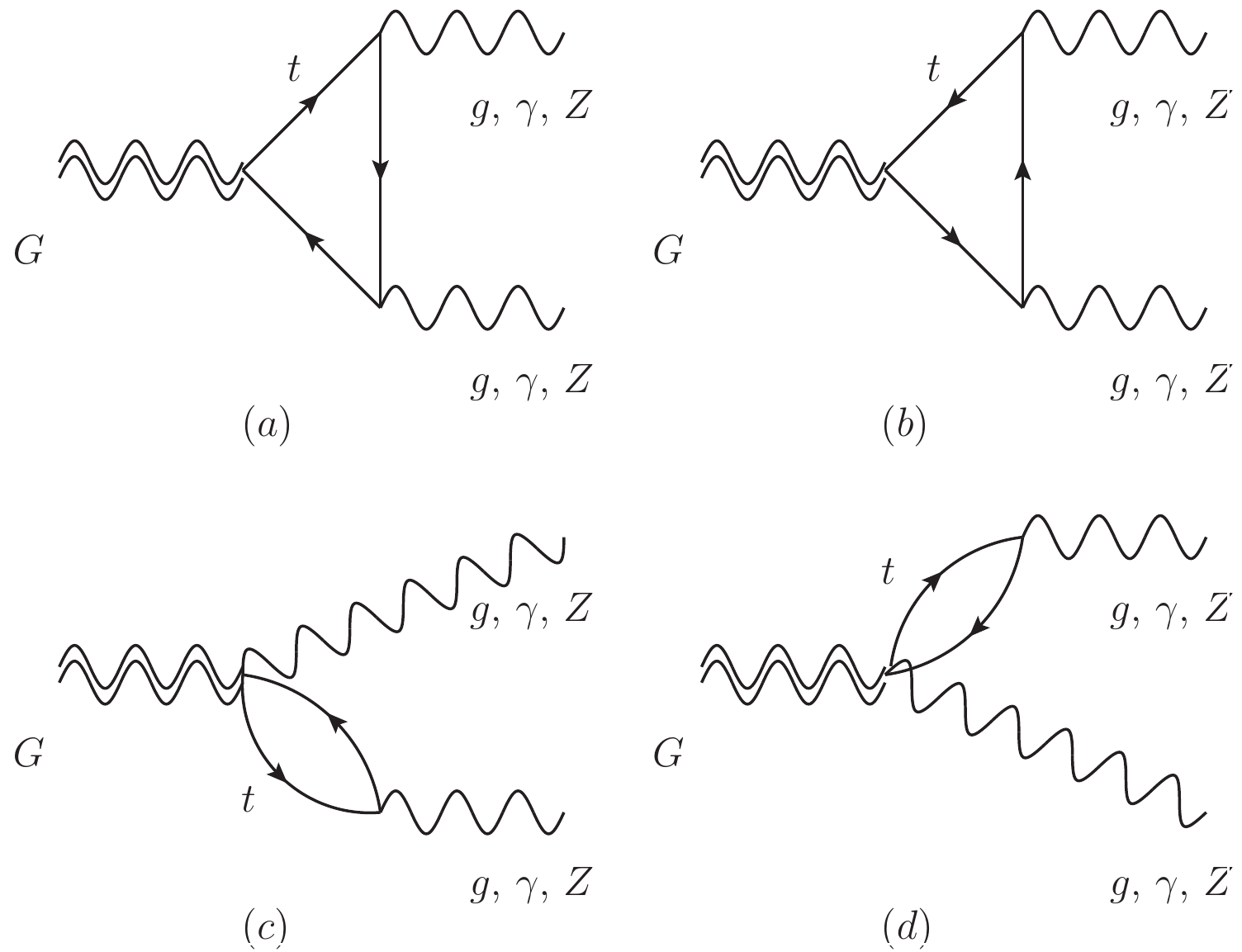}
\caption{Feynman diagrams for the one-loop top-quark contributions to the $Ggg$, $G\gamma\gamma$, $G\gamma Z$, and $GZZ$ vertices.
}
\label{fig:FeynD}
\end{figure}

Before proceeding to discuss the LHC phenomenology of the KK graviton $G$, let us begin by discussing what the implication is for the power counting rules in Eq.~(\ref{PowerCounting}). By assumption, the  right-handed top quark coupling $c_{tt}$ is of ${\cal O}(1)$, while the couplings with $g$, $\gamma$ and $Z$ are suppressed by the extra-dimensional volume dilution with the order of $\alpha_{(s)}/(4\pi)$. Thus, it is expected that the top-quark one-loop contributions to the couplings between the KK graviton $G$ and these SM gauge bosons  as shown in Fig.~\ref{fig:FeynD} should be at the same order as the tree-level ones. 
In other words, the leading-order (LO) interactions of $G$ with $g$, $\gamma$ and $Z$ should be the combination of these two contributions. Therefore, it is useful to define the following LO effective couplings, given by
\begin{eqnarray}\label{Ceff}
c^{\rm eff}_{gg} &=& \left\{ \begin{array}{cc}
c_{gg}(m_G) + {c_{tt}}A_G\left(\frac{4m_t^2}{m_G^2}\right) \,, & m_G \geq m_t \\
c_{gg}(m_t) + c_{tt} B_G\left(\frac{4m_t^2}{m_G^2}\right)\,, & m_G < m_t \\
\end{array}
\right.\nonumber\\
c^{\rm eff}_{\gamma\gamma} &=& \left\{ \begin{array}{cc}
c_{\gamma\gamma}(m_G) + 2 Q_t^2 N_c {c_{tt}}A_G\left(\frac{4m_t^2}{m_G^2}\right) \,, & m_G \geq m_t \\
c_{\gamma\gamma}(m_t) + 2Q_t^2 N_c c_{tt} B_G\left(\frac{4m_t^2}{m_G^2}\right)\,, & m_G < m_t \\
\end{array}
\right.
\end{eqnarray}
where $Q_t = 2/3$, $N_c = 3$ and $m_t = 173.1$~GeV represent the top quark electric charge, color and mass, respectively. 
We have also defined the following functions
\begin{eqnarray}
\begin{array}{cc}
A_G(\tau) \equiv -\frac{1}{36}\left[9\tau(\tau+2)f(\tau) +6(5\tau+4)g(\tau) - 39\tau -35 +12\ln(\tau/4)\right]\,, & \tau \leq 4 \\
B_G(\tau) \equiv -\frac{1}{36}\left[9\tau(\tau+2)f(\tau) +6(5\tau+4)g(\tau) - 39\tau  -35 \right]\,, & \tau > 4 \\
\end{array}
\end{eqnarray}
with
\begin{eqnarray}
f(\tau) &=&\left\{\begin{array}{cc}
 -[{\rm arctanh}(\sqrt{1-\tau})-i\pi/2]^2\,, & \tau < 1 \\
 \arcsin^2(1/\sqrt{\tau})\,,& \tau \geq 1 \\
\end{array}\right. \\
g(\tau) &=& \left\{ \begin{array}{cc}
\sqrt{1-\tau}[{\rm arctanh}(\sqrt{1-\tau})-i\pi/2]\,,& \tau < 1 \\
\sqrt{\tau-1} \arcsin(1/\sqrt{\tau})\,,& \tau \geq 1
\end{array}
\right.\,.
\end{eqnarray}
Note that in order to keep the gauge invariance of the right-handed-top-quark-$G$ coupling, we have included the coupling of
$\bar{t}_R G^{a\,(\mu} \gamma^{\nu)} t_R G_{\mu\nu}$  in the Lagrangian of Eq.~(\ref{LagB}) by the covariant derivative of $t_R$. 
As a result, we need to incorporate the one-loop Feynman diagrams (c) and (d) of Fig.~\ref{fig:FeynD} besides the triangle ones calculated in Ref.~\cite{Geng:2016xin}. 
However, as shown in Appendix~\ref{AddD}, their contributions vanish identically when we apply the KK graviton and EW boson on-shell conditions.  

\begin{figure}[ht]\center
 \includegraphics[width=0.8\textwidth]{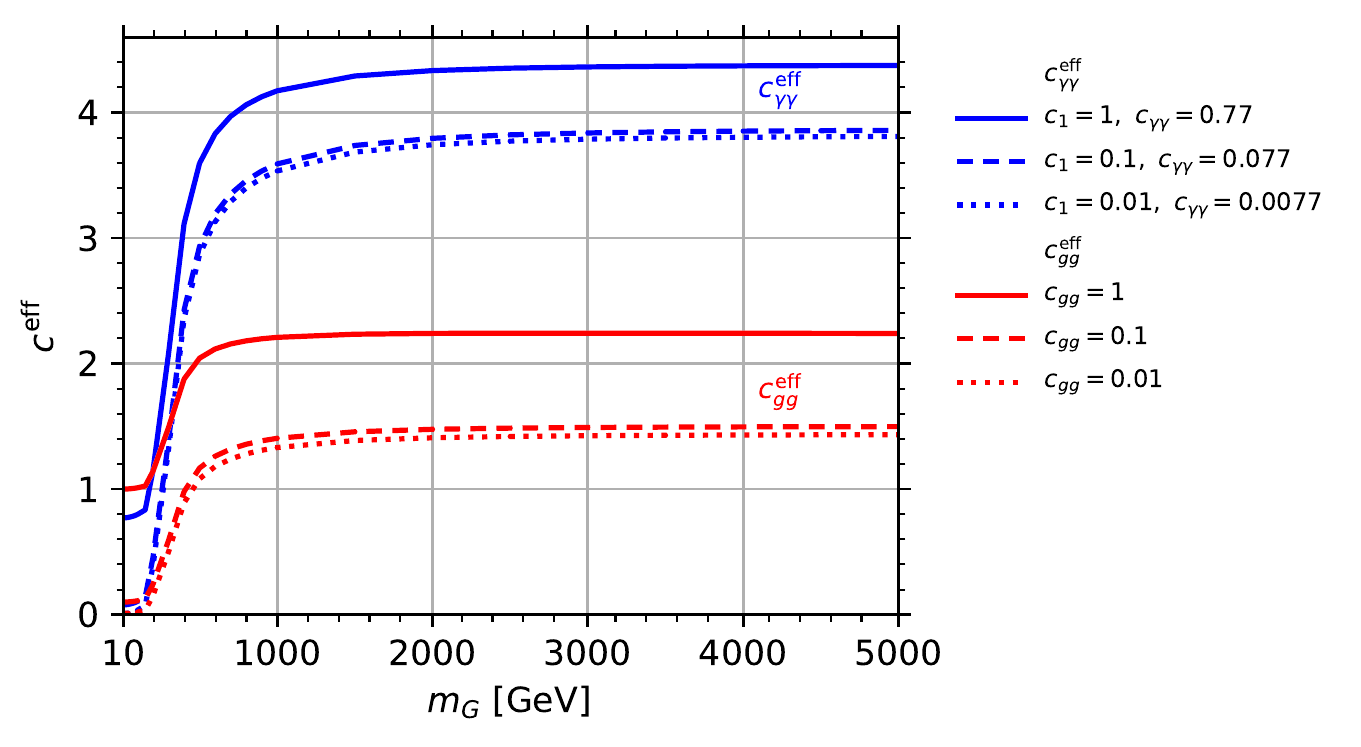}
\caption{Effective KK graviton couplings as  functions of the KK graviton mass,
 where $c_{tt} =1$ is chosen and the solid, dashed and dotted lines correspond to $c_1$ and $c_{gg}$= 1, 0.1 and 0.01, while
 the blue and red lines are for the effective coupling between the KK graviton and photons and gluons, respectively.
}
\label{fig:eff_coupling}
\end{figure}

Compared the top-loop contributions to the $Ggg$ and $G\gamma\gamma$ couplings in Ref.~\cite{Geng:2016xin}, 
we have extended the valid range of the expressions to the whole parameter space, no matter  whether $m_G$ is larger than $2m_t$ or not. 
In addition, the top-quark one-loop diagrams in Fig.~\ref{fig:FeynD} are UV divergent, so that we need to perform the renormalization skill to remove the corresponding UV divergence, which results in the renormalization scale  dependence when defining the tree-level couplings $c_{gg(\gamma\gamma)}$ and the modification of the loop functions from $A_G$ to $B_G$ when $m_G$ decreases below $m_t$. 
Concretely, when $m_G > m_t$, the appropriate renormalization scale should be $m_G$ since the KK graviton is on-shell at this scale. 
In comparison, if $m_G < m_t$, we need to integrate out the top quark fields first in the theory so that the renormalization scale should be chosen 
to be $m_t$. Nevertheless, the final effective coupling $c_{gg(\gamma\gamma)}^{\rm eff}$ is a continuous function of $4m_t^2/m_G^2$, whereas its first derivative is not, which is the reflection of the renormalization effects. A further justification of the expression in Eq.~(\ref{Ceff}) is provided by looking at the so-called decoupling limit in which $m_t/m_G \to \infty$. It is easy to check that in this limit the loop function $B_G(4m_t^2/m_G^2)$ vanishes, which agrees with the decoupling theorem. In Fig.~\ref{fig:eff_coupling}, we show the typical behavior of the effective couplings for $G\gamma\gamma$ and $Ggg$ as functions of the KK graviton mass $m_G$, which clearly displays the decoupling tendency of top-loop effect when $G$ becomes lighter.

\section{Production Cross Section of the KK Graviton}\label{sec:production}
\begin{figure*}\center
 \includegraphics[width=0.8\textwidth]{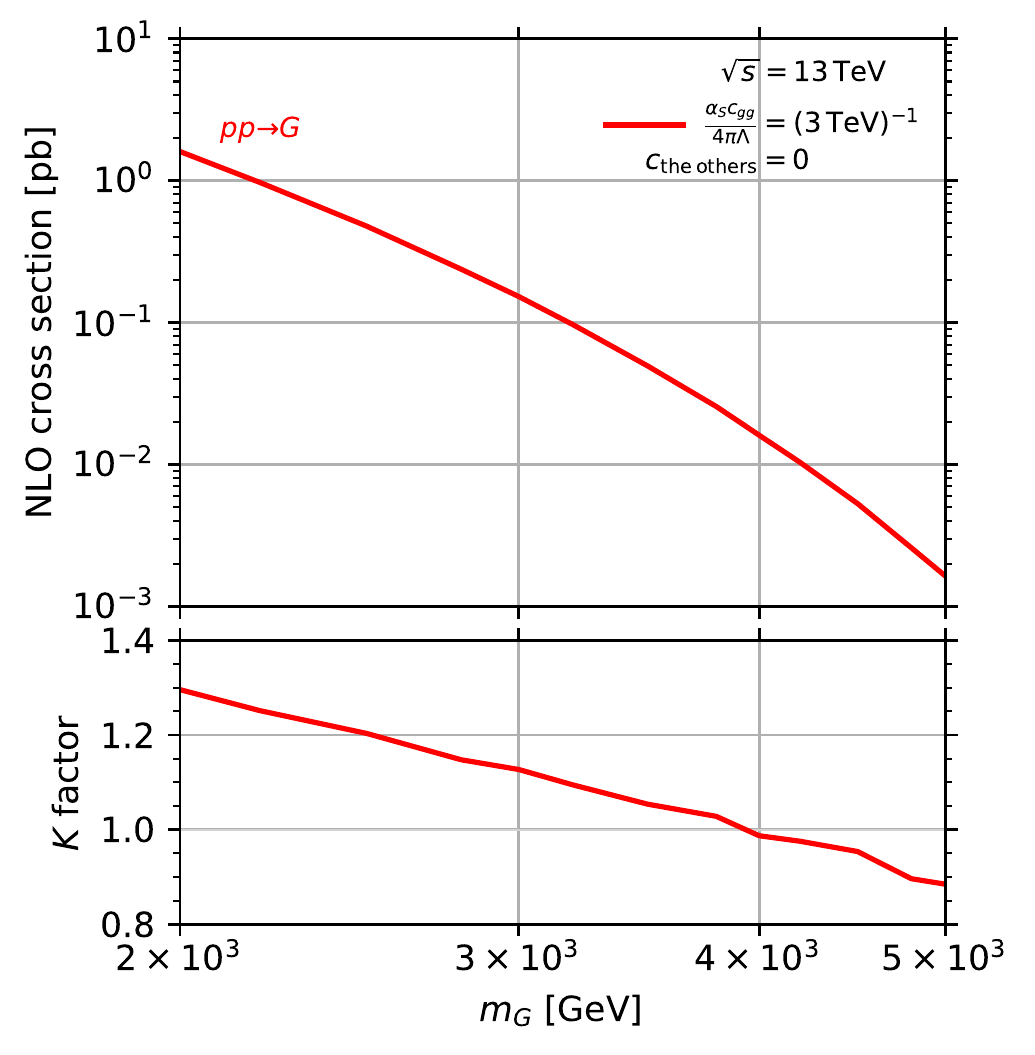}
\caption{Production cross sections of the KK graviton with the NLO accuracy at the 13~TeV LHC (upper panel)
and $K$ factors (lower panel)
as  functions of the KK graviton mass, where
$\alpha_{s} c_{gg}/(4\pi\Lambda) = (3~\mathrm{TeV})^{-1}$ and the KK graviton couplings to the others including the top quark are zero.
}
\label{fig:kk_xsec_nlo}
\end{figure*}
In our model, the inclusive KK graviton production of $pp \to G$ at the LHC is given by
a tree level contribution to $gg \to G$ as well as those via right-handed top loops
denoted as ``LO'' (at the same order).
 For the estimation of ``NLO'' contributions, both 1 and 2-loop level calculations are required
because the right-handed top loops already exist at ``LO''.
For a rough estimation with the NLO QCD accuracy, we depict $pp \to G$ for the $p p$ collisions at 13 TeV as the function of the KK graviton mass 
 in the range of 2--5~TeV in Fig.~\ref{fig:kk_xsec_nlo}, where
 we have assumed that gluons can only interact with the KK graviton.
In our numerical analysis, we use {\sc Madgraph5\_aMC@NLO}~\cite{Alwall:2014hca,Mastrolia:2012bu,Peraro:2014cba,Hirschi:2016mdz}
with the LO/NLO {\sc NNPDF2.3}~\cite{Ball:2012cx}. 
We also take $\alpha_{s} c_{gg}/(4\pi\Lambda) = (3~\mathrm{TeV})^{-1}$, which does not affect  the $K$ factors. Note that the values of $K$ 
 are smaller than 1 at the high mass region when the KK graviton only couples to a gluon current.
For the KK graviton mass range considered, the $K$ factors are within $30~\%$ discrepancy from 1,  corresponding to a tree-level production.
 Since these factors become larger as the KK graviton mass  decreases,  in this study we only concentrate on 
 the high mass range of $m_G=2$--$5$~TeV.

\begin{figure*}\center
 \includegraphics[width=0.8\textwidth]{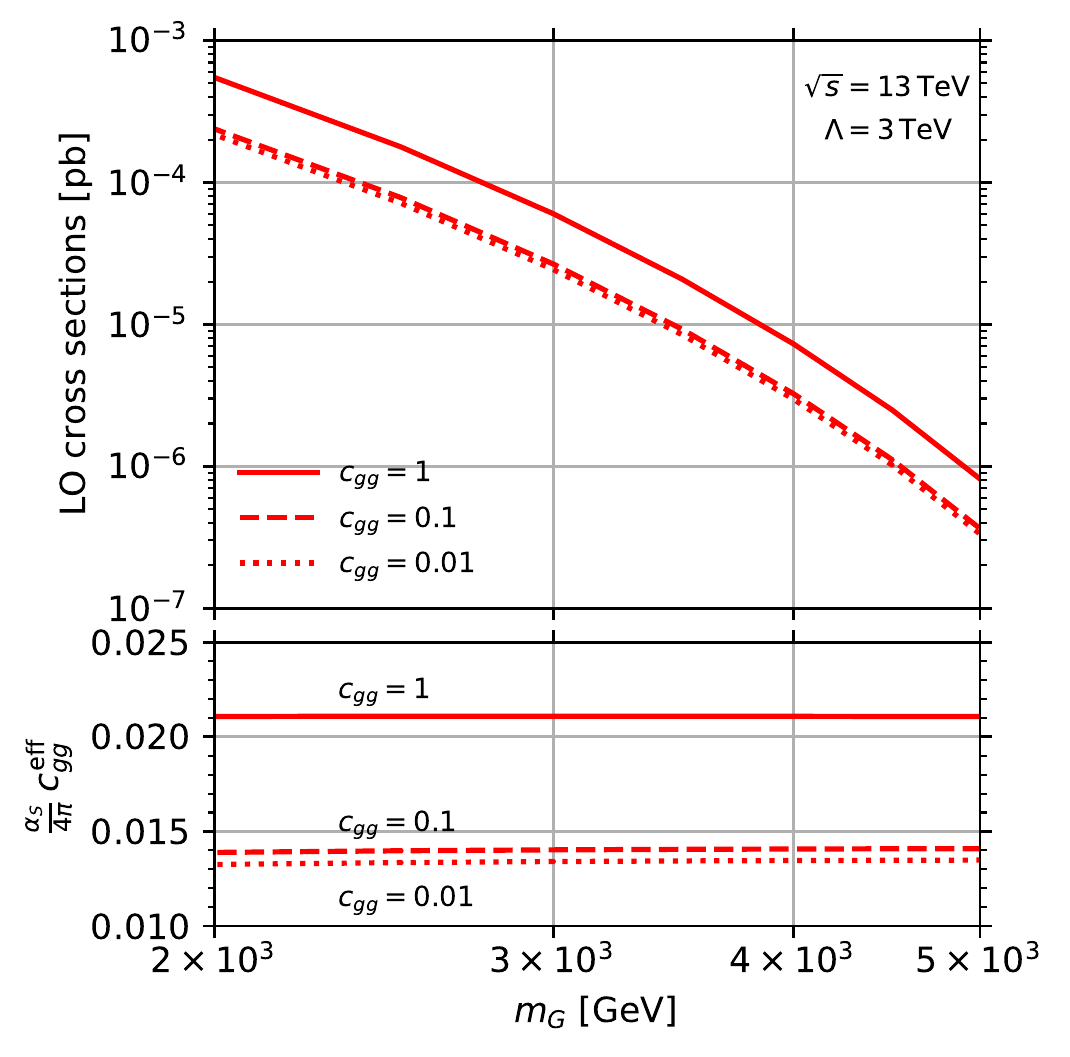}
\caption{
Production cross sections of the KK graviton with the ``LO'' accuracy (upper panel), which includes the production via the right-handed top loop
at the 13~TeV LHC, and $\alpha_{s} c^{\mathrm{eff}}_{gg}/(4\pi\Lambda)$ (lower panel)
 as functions of the KK graviton mass, where 
 $\Lambda = 3~\mathrm{TeV}$ and $c_{tt} =1$, along with
  three choices of $c_{gg}=1$  (solid),
 $0.1$ (dashed)  and $0.01$ (dotted).
}
\label{fig:kk_xsec_lo}
\end{figure*}
 As these NLO cross sections are rough estimations,
 we use the  ``LO'' cross section, which includes the production via the right-handed top  loops in our calculation.
For simplicity, we take $c_{tt}=1$ in the whole analysis.
The results  of the  ``LO'' cross section as a function of KK graviton mass are shown in Fig.~\ref{fig:kk_xsec_lo}.
The effective coupling between the KK graviton and a gluon current, $(4\pi)^{-1} \alpha_Sc^{\mathrm{eff}}_{gg}$, is
 shown in the lower panel.
Because of the high KK graviton mass, the top-loop effect is almost the same as that
in Fig.~\ref{fig:eff_coupling} with $m_G=2$--$5$~TeV.
In addition, the lines for $\alpha_{s} c^{\mathrm{eff}}_{gg}/(4\pi\Lambda)$  at $c_{gg}=0.1$ and $c_{gg}=0.01$ are in the same order because of 
the large contribution from the top-loop.
In the following sections, we will use $c_{gg}=1$ in our analysis as it is a natural value 
from a profile view of the gluon field in a bulk. 
The top-loop contributions are the same order as those in the tree-level as shown in Fig.~\ref{fig:eff_coupling}.

\section{Decays of the KK Graviton}\label{sec:decay}

\begin{figure*}\center
 \includegraphics[width=0.8\textwidth]{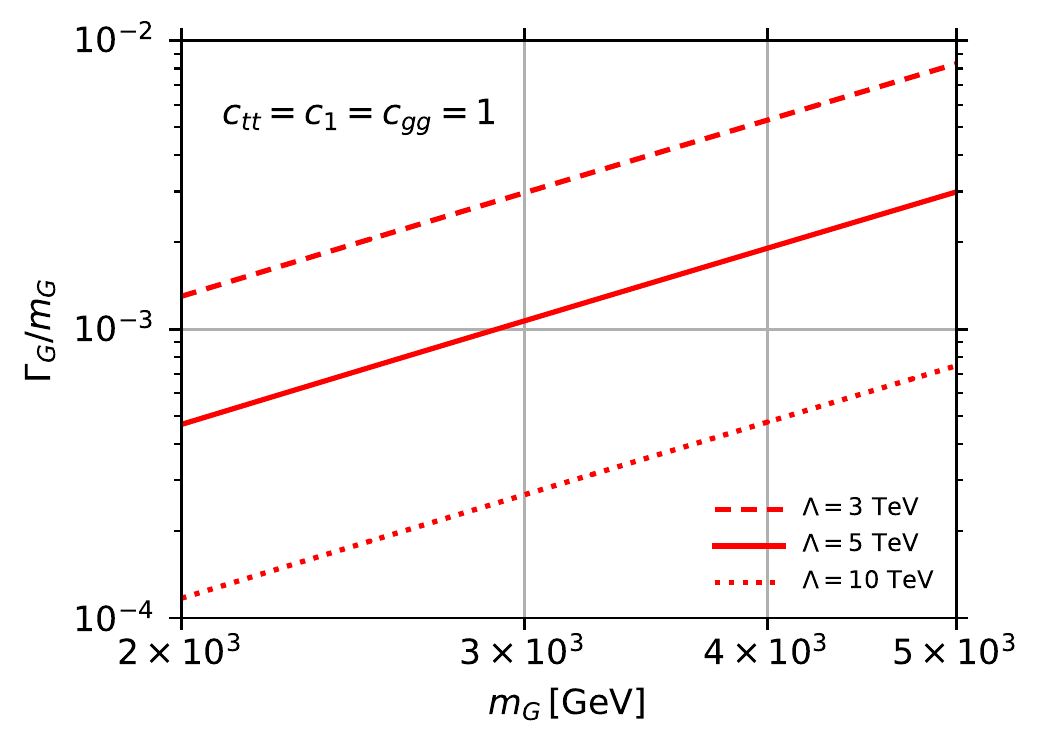}
\caption{Total width of the KK graviton  to its mass as a function of the KK graviton mass,
 where $c_{tt}=c_{1}=c_{gg} =1$  with $\Lambda=3$ (dashed), $5$ (solid) and $10$ (dotted) TeV.
}
\label{fig:width}
\end{figure*}

 Our signals are  
$t\bar{t}$, $gg$, $\gamma\gamma$, $\gamma Z$ and $ZZ$
 through the decays of the KK graviton resonance.
We assume that the narrow width approximation can be applied for the  relativistic Breit-Wigner resonance of the KK graviton.
In this case, the cross sections of the signals are obtained by
\begin{equation}
\sigma(pp \to G \to XX') = \sigma(pp \to G) \mathrm{B}(G \to XX'),
\end{equation}
where $\sigma(pp \to G)$ is the production cross section of the KK graviton and
$\mathrm{B}(G \to XX')$ correspond to the branching ratios of the KK graviton decaying into the particle pairs of $XX'$ with
$XX'= t\bar{t}$, $gg$, $\gamma\gamma$, $\gamma Z$, and $ZZ$.
Figure~\ref{fig:width} shows the KK graviton ($G$) total width divided by its mass, $\Gamma_G/m_G$.
Note that {\sc MadWidth}~\cite{Alwall:2014bza} provides the partial decay 
rates numerically for each mass point.
The resonance is very narrow as $\Gamma_G/m_G < 1 \%$ in a range of 2--5 TeV of $m_G$ with bench mark points of 
$\Lambda=3, 5, 10$~TeV. respectively.
 
\begin{table}[t]\center
\caption{Branching ratios of the KK graviton for $c_{tt} = c_{1} = c_{gg} =1$.\newline}
\begin{tabular}{c|c c c c c}
\hline 
 & \multicolumn{5}{c}{Branching ratios $[\%]$} \\
  $m_G [{\rm GeV}] $	& $t\bar{t}$ & $gg$ & $\gamma\gamma$ & $\gamma Z$ & $ZZ$ \\
\hline
\hline
2000   & &  &   &              &  \\
--   	   & $\sim100$\,\,\,\,\, &  $0.5$\,\,\,\,\, & $0.001$\,\,\,\,\, & $6 \times 10^{-4}\,\,\,\,\,$  & $1 \times  10^{-4}\,\,\,\,\,$       \\
5000   &    		  &            &              				 &          & \\
\hline
\end{tabular}
\label{tab:br}
\end{table}
The partial widths for $G \to XX'$
are given by
\begin{eqnarray}
\Gamma(G\to t\bar{t}) &=& \frac{N_c}{320\pi} \frac{c_{tt}^2 m_G^3}{\Lambda^2} \left(1-\frac{4m_t^2}{m^2_G}\right)^{3/2} \left(1+\frac{8}{3} \frac{m_t^2}{m_G^2}\right)\,, \label{ttRate} \\
\Gamma(G\to \gamma\gamma) &=& \frac{m_G^3}{80\pi\Lambda^2} \left|\frac{\alpha}{4\pi} c_{\gamma\gamma}^{\rm eff}  \right|^2\,, \\
\Gamma(G\to gg) &=& \frac{m_G^3}{10\pi\Lambda^2} \left|\frac{\alpha_s c_{gg}^{\rm eff}}{4\pi}\right|^2\,,\\
\Gamma(G\to \gamma Z) &=& \frac{m_G^3}{160\pi\Lambda^2} \left(\frac{\sin 2\theta_W}{\cos^2 \theta_W}\right)^2 \left(1-\frac{m_Z^2}{m_G^2}\right)^3 \\ \nonumber
	&\times& \left(1+ \frac{m_Z^2}{2m_G^2} + \frac{m_Z^4}{6m_G^4}\right)  \left|\frac{\alpha c_{\gamma\gamma}^{\rm eff}}{4\pi} \right|^2 \, \label{ZpRate} , \\
\Gamma(G\to ZZ) &=& \frac{m_G^3}{80\pi\Lambda^2} \tan^4\theta_W \left(1-\frac{4 m_Z^2}{m_G^2}\right)^{1/2} \\ \nonumber
	&\times& \left(1 - \frac{3 m_Z^2}{m_G^2} + \frac{6 m_Z^4}{m_G^4}\right) \left|\frac{\alpha c_{\gamma\gamma}^{\rm eff}}{4\pi} \right|^2 \,. \label{ZZRate}
\end{eqnarray}
Note that $\Gamma(G\to t\bar{t})$
 contains  an extra  factor of $1/2$ compared to {that when the KK graviton $G$ couples to both the left- and right-handed top quarks}. 
The decay mode with $\gamma Z$  in our model  appears due to
the non-universal couplings  of $G$ to the weak gauge bosons.
Since in the following we only consider the heavy KK graviton case in which $m_G \gg m_Z$, we can take the zero $Z$ mass limit when computing the one-loop corrected effective couplings, which are reduced to $c^{\rm eff}_{\gamma\gamma}$ as in Eqs.~(\ref{ZpRate}) and (\ref{ZZRate}).     

As shown in Table~\ref{tab:br}, the $t\bar{t}$ channel has almost $100~$\% branching ratio for the KK graviton mass in 2--5~TeV.
The other channels have small fractions of the branching ratios,
 which can still lead to some constraints on our model parameters
as a diphoton final state is a clean experimental signature.
The branching ratios of  $gg$,  $\gamma\gamma$, $\gamma Z$ and $ZZ$  channels are $0.5~\%$,  $0.001~\%$,
 $6\times 10^{-4}~\%$, and  $1\times 10^{-4}~\%$, respectively.
As we concentrate on a high mass region of the KK graviton,
all decay modes in our model  are kinematically allowed 
and the corresponding branching ratios are almost fixed in the whole mass range.\\

\section{Constraints from the 13~TeV LHC data}\label{sec:limit}
\begin{table*}[t!]\center
\caption{Constraints from resonance searches based on $\sqrt{s} = 13~$TeV at the LHC with
the observed 95\% CL upper limits on the resonant production cross section ($\sigma$) $\times$ branching ratio ($B$) ($\times$ acceptance ($A$)).} 
\scalebox{0.7}{
\begin{tabular}{l|l|l|l|l}
\hline
 Decay mode & Reference & Limit Table/Figure & Limit on & $L$\,(fb$^{-1}$)  \\ 
\hline\hline
 $t\bar{t}$ & CMS-PAS-B2G-17-017~\cite{CMS:2018rzm} & Table 2 
 & $\sigma({\rm narrow-width}~Z')\times B$ & 36 \\
 \hline
 $gg$ & CMS arXiv:1806.00843~\cite{Sirunyan:2018xlo} & Table 2 (gg) 
 & $\sigma({\rm narrow-width~Res.})\times B\times A$ & 27+36 \\
 \hline
 $\gamma\gamma$ & CMS-PAS-EXO-17-017 \cite{CMS:2018thv} & Fig. 3 (middle) 
 & $\sigma(G_{\rm RS})\times B$ & 35.9 \\
\hline
 $\gamma Z$ & CMS-PAS-EXO-17-005 \cite{CMS:2017fge} & Fig. 6a
 & $\sigma({\rm narrow-width~Res.}) \times B$ & 35.9 \\
\hline
 $ZZ$ & ATLAS-CONF-2018-016 \cite{ATLAS:2018tpf} & Fig. 14b 
 & $\sigma(G_{\rm RS})\times B$ & 79.8 \\
 \hline
\end{tabular}
}
\label{tab:rsearch}
\end{table*}

Our target signals  come from the decay of the KK graviton resonance in the $s$-channel.
The $\sqrt{s} = 13~$TeV LHC results are taken to limit the model in the high mass region of the KK graviton.
We take $c_{tt} = c_1 = c_{gg} = 1$ in Eq.~(\ref{LagB}) in our analysis.
We choose the CMS data for $t\bar{t}$, $gg$ (dijet),  $\gamma\gamma$, $\gamma Z$~\cite{CMS:2018rzm,Sirunyan:2018xlo,CMS:2018thv,CMS:2017fge}
and ATLAS data for $ZZ$~\cite{ATLAS:2018tpf} final states.
In Table~\ref{tab:rsearch}, we list the current results for the resonance searches,  which are used to constrain  the model parameter space of
the top-philic bulk RS model. 
 In particular, the constraint on a $t\bar{t}$ resonance is given in terms of the narrow-width $Z^\prime$ search in Ref.~\cite{CMS:2018rzm},
while the bounds on the model independent narrow-width resonances are studied for 
$gg$ (dijet) and $\gamma Z$ channels in Ref.~\cite{Sirunyan:2018xlo,CMS:2017fge}.
In addition, we use the gluon-gluon ($gg$) resonance result for our dijet analysis as the dominant contribution stems from the gluon fusion.
In Refs.~\cite{CMS:2018thv, ATLAS:2018tpf}, the RS graviton is considered for $\gamma \gamma$ and $ZZ$ modes.
For the $gg$ (dijet) mode, a limit can be given for the fiducial cross section of $\sigma(pp \to G) \times \mathrm{B} \times A$, where 
$A$ is acceptance. We apply the fiducial cuts at the parton level 
to obtain the fiducial cross section.
The other limits for  $\sigma(pp \to G) \times \mathrm{B}$ are also available as shown in the column  of ``Limit on'' in 
Table~\ref{tab:rsearch}.
 The production cross sections of the
KK graviton and its decay branching ratios have been discussed in Secs.~\ref{sec:production} and \ref{sec:decay}, respectively.
We note that we extract the data from the figures by using {\sc WebPlotDigitizer}~\cite{WebPlotDigitizer} for $\gamma\gamma$, $\gamma Z$, and $ZZ$ modes.

\begin{figure*}\center 
 \includegraphics[width=0.9\textwidth]{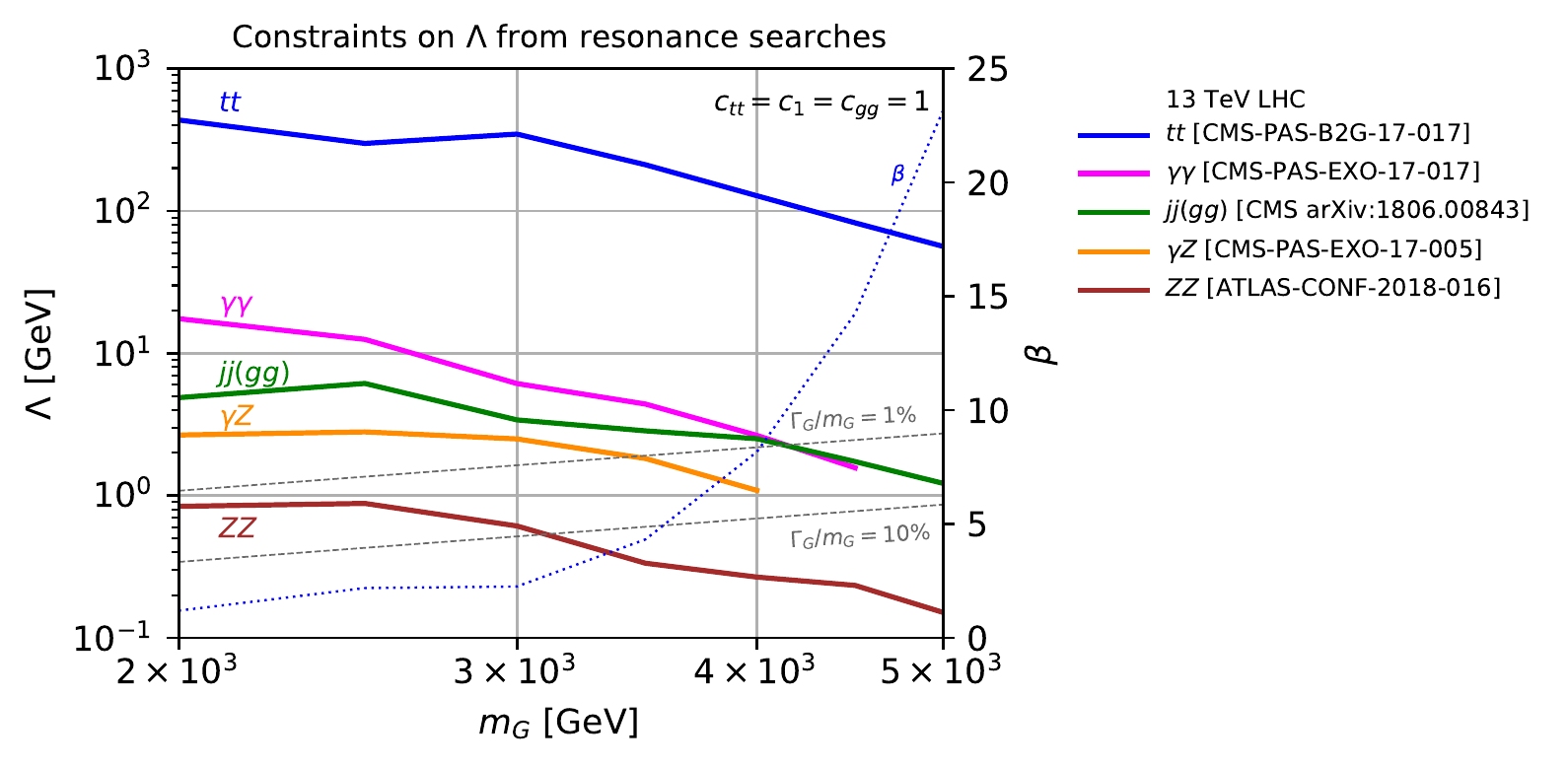} 
\caption{Constraints on $\Lambda$ as a function of the KK graviton mass from the observed 95\% CL lower limits of 
the resonance searches at the 13~TeV LHC with the regions below each line excluded,
where the grey dashed and blue dotted lines correspond to the KK graviton width-to-mass ratio and $\beta=k/\bar{M}_\mathrm{Pl}$, 
respectively.
}
\label{fig:exclusion_resonances}
\end{figure*}

Figure.~\ref{fig:exclusion_resonances} shows the 
constraints on the inverted coupling $\Lambda$ from 
the observed $95~\%$ CL lower limits of the resonance searches listed in Table~\ref{tab:rsearch}.
The $t\bar{t}$ final state of the KK graviton resonance signal gives the  strongest limit
as expected from  its almost $100~\%$ branching ratio.
Although the branching fraction of $0.001~\%$ for the $\gamma\gamma$ channel is small,
it leads to the next strongest limit due to its 
 clean signature in the experiment.
 The dijet signal ($gg$ resonance), which provides the second largest branching ratio of $G$ decays, yields the comparable result for $m_G\sim 4$--$4.5~$TeV but 
a weaker limit for $m_G < 4~$TeV than $\gamma\gamma$ 
because the acceptance after cuts is small.
For the KK graviton production and its decay mode of  $gg$ at the parton level,
the efficiency is about $0.3$ for the 2--5 TeV mass after imposing  $|\eta_{j}| < 2.5$ and $|\Delta \eta_j| < 1.3$, 
where $\eta_{j}$ is  the pseudorapidity of each jet.
It is clear that the cut of $|\Delta \eta_j| < 1.3$ abandons many signal events.
For example, for $m_G=2$~TeV, $67 \%$ signal events are excluded with $|\Delta \eta_j| < 1.3$ 
after imposing $|\eta_{j}| < 2.5$ beforehand.
Because the  $jj$ background is from $t$ and $u$ channels besides the $s$ one,
forward and backward regions should be cut to reduce the background events.
Our signal is a resonance ($s$-channel), the central region has relatively more events  than the background ones.
Note that  the structure of matrix elements is important. 
On the other hand, the angular momentum (i.e., $s$ or $d$-wave) is essential for the angular distribution,
which can be used  to  distinguish spins of the  signal resonances.
We  now  concentrate on the total cross sections.
As seen in Fig.~\ref{fig:exclusion_resonances}, the strong signals are given by 
the $t\bar{t}$, $jj (gg)$, $\gamma Z$, and $ZZ$ modes.
In the figure, we show $\Gamma_G/m_G = 1 \%$ and $10 \%$ points with grey dashed lines
as we assume a narrow-width in our calculations of the signals and 
 use the information of the narrow-width resonances from the experimental data.
Lower limits on $\Lambda$ for each mass point in our model are less than $500$~GeV,
 which are different from those of several TeV or several 10~TeV on $\Lambda$ in the universal case
with $m_G$ in the range of $2$--$5$~TeV~\cite{Kraml:2017atm}.
 In order to relate our phenomenological parameters $m_G$ and $\Lambda$ to the bulk geometry more transparently, 
 we have defined a new variable $\beta \equiv (3.83)^{-1} \times m_G/\Lambda = k/\bar{M}_\mathrm{Pl}$ and shown its upper limits as the dashed curve at each $G$ mass in Fig.\ref{fig:exclusion_resonances}. As a result, the values of $\beta \geq 0.01$ on the curve imply large 5D curvatures $k$,
which cannot be obtained by a simple assumption
in the string theory~\cite{Davoudiasl:1999jd}. Concretely, the corresponding radii of the extra dimension should vary from $9.6/\bar{M}_\mathrm{Pl}$ at $m_G = 2$~TeV 
to $0.53/\bar{M}_\mathrm{Pl}$ when $m_G = 5$~TeV.

%
\begin{figure*}[t]\center
 \includegraphics[width=0.8\textwidth]{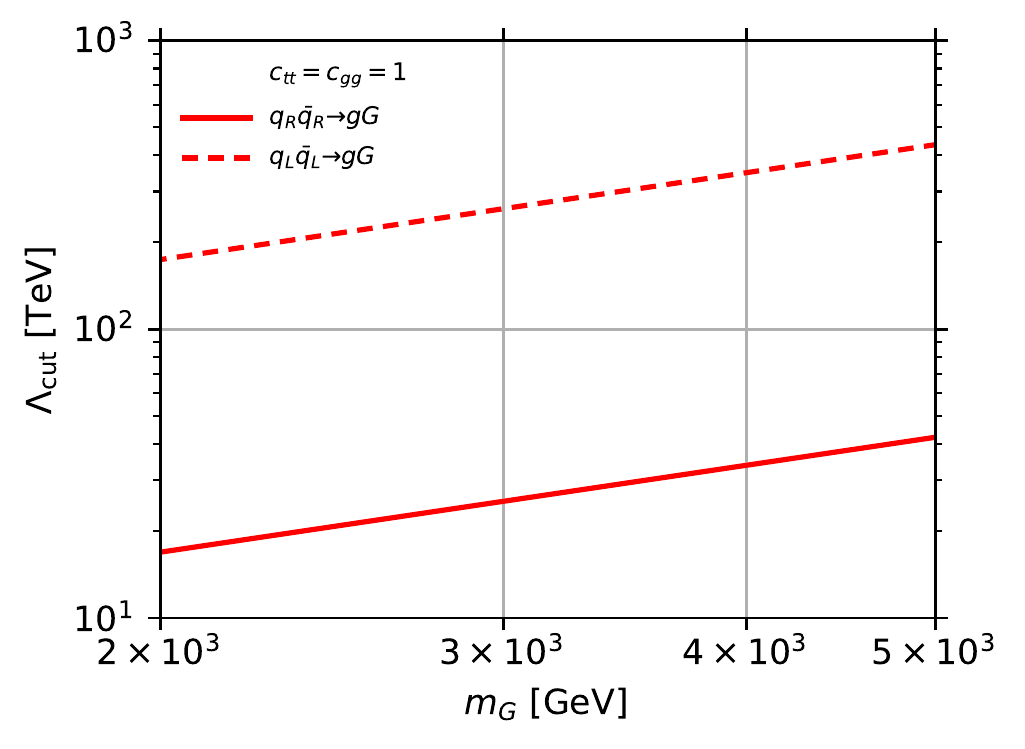} 
\caption{The cutoff scale $\Lambda_\mathrm{cut}$ as a function of the KK graviton mass  
with the region bellow each line allowed.}
\label{fig:et_cutoff}
\end{figure*}

 We now examine the largest allowed cutoff scale $\Lambda_{\rm cut}$ in our model, which can be derived from the violation of the perturbative unitarity.
It is well-known that $\Lambda_{\rm cut}$ becomes lower for the non-universal coupling case~\cite{Artoisenet:2013puc, Falkowski:2016glr}. In the present model, the strongest constraints can be obtained from the $s$-wave processes of $t_R \bar{t}_R \to g G$ and $q_L \bar{q}_L \to g G$,
in which the 0-helicity KK graviton production can generate the amplitudes proportional to $[(c_g - c_{tR})$ or $(c_g - c_{qL})]$ $\times s^{3/2}/(m^2_G \Lambda)$~\cite{Falkowski:2016glr},
where the subscripts $R$ and $L$ denote the projection operators of
$(1\pm\gamma_5)/2$,  respectively.
We utilize the formula in Ref.~\cite{Falkowski:2016glr} and depict $\Lambda_\mathrm{cut}$ in Fig.~\ref{fig:et_cutoff},
where the perturbative unitarity is lost  when $\mathrm{Re}(\mathrm{amplitude}) > 1/2$.
Figure~\ref{fig:et_cutoff} shows that the $t_R \bar{t}_R \to g G$ process gives a stronger limit on  
the cutoff  of $\Lambda_\mathrm{cut}$ for each mass point than the $q_L \bar{q}_L \to g G$ one.
 The values of $\Lambda_\mathrm{cut}$ can be of several 10~TeV,
which are much higher than the experimental lower limit of $\Lambda < 500$~GeV given in
Fig.~\ref{fig:exclusion_resonances}.

\section{Summary}\label{sec:summary}
We have concentrated on the  bulk RS model, in which 
 the KK graviton  $G$ interacts strongly with the right-handed top quark due to
the profile of the right-handed top quark localized near the IR brane.
In our model, as the color $SU(3)_c$ and hypercharge $U(1)_Y$ gauge fields  propagate in the bulk,  
the corresponding couplings with the KK gravitons are suppressed by a volume factor. 
 In contrast, the other SM particle fields are localized near the UV brane, which give exponentially small couplings of these particles to $G$. 
We have studied the constraints on the model parameter space based on the current 13~TeV LHC results for the first/lightest
KK graviton.
Our LHC signals are $t\bar{t}$, $\gamma\gamma$, $jj (gg)$, $\gamma Z$, and $ZZ$ via the decays of the KK graviton resonance.
In our estimations, we have  included the top-loops for the production of the KK graviton and its decays.
We have found that the strongest limit is from the $t\bar{t}$ mode, which gives  $\mathcal{O}(100)$~GeV for the
lower bounds of $\Lambda$ for $m_G=2$--$5$~TeV, which are smaller than those in the universal coupling case.
Our rough  unitarity bounds in the model suggest that $\Lambda$ should be smaller than several 10~TeV for $m_G=2$--$5$~TeV.
A parameter region in our non-universal model is survived roughly between $ \mathcal{O}(100)$~GeV$ < \Lambda <$ several 10~TeV
for $m_G=2$--$5$~TeV.

 Finally, we would like to remark that one interesting prediction of our present scenario is the presence of light KK states of bulk gauge bosons, such as massive KK gluons. When the bulk geometry is a slice of AdS$_5$ spacetime with its curvature (length) represented as $k$ ($L$), the mass of the first massive KK gluon $g^1$ is found to be $m_{g^1} \approx 2.4 k e^{-kL}$, and its coupling to the right-handed top quark is $g_{g^1} \sim g_s (kL)^{1/2}$. When $kL\sim 30$, it is predicted that this mass of this KK gluon is of ${\cal O}$(TeV), and  more strongly coupled to quarks than the usual gluon, which is promising to be probed at colliders. In fact, this massive KK gluon $g^1$ is even parametrically lighter than the first KK state of graviton whose mass is $m_G \approx 3.8 k e^{-kL}$. Thus, $g^1$ is expected to be more easily discovered  than $G$, which violates our implicit assumptions that the KK graviton $G$ is the first KK state to be seen at the LHC. One way to avoid this problem is to find a way to make the KK graviton $G$ lighter than the KK gluon $g^1$ 
It is shown in Ref.~\cite{Falkowski:2016glr} that we can achieve this by adding the boundary kinetic terms for the bulk graviton at both UV and IR branes.

It is interesting to compare the phenomenology of this top-philic KK gluon~\cite{Lillie:2007yh} with that of the top-philic KK graviton. In the light of the Landau-Yang theorem, the massive KK graviton cannot be singly produced on-shell via the gluon fusion, neither can it decay to the diphoton or digluon final states. Unlike the colorless top-philic vector boson studied in Ref.~\cite{Greiner:2014qna}, the KK gluon here carries color quantum numbers. Thus, it is shown in Ref.~\cite{Lillie:2007yh} that, due to its nonvanishing constant 5D field profile in the IR, $g^1$ can still have sizable couplings to the light quarks, 
leading to
 that $g^1$ is mostly produced on-shell by the $q\bar{q} \to g^1$ with $q$ denoting light quarks contained in the proton. On the other hand, we expect that 
 the gluon fusion can also give rise to the substantial $g^1$ production at the LHC, with the channels as $gg \to g^1 g $ in which $g_1$ decays to a $t\bar{t}$ pair, or $gg \to g^1 \to t\bar{t}$, where $g^1$ is created off-shell~\cite{Greiner:2014qna}. However, in contrast to the case in Ref.~\cite{Greiner:2014qna}, we can prove by some simple estimations that both processes are dominated by tree-level diagrams. Take, for instance, the $g^1$ on-shell production  associated with a jet from the gluon fusion. At the tree level, the amplitude should be of order $\sim g_s^2$. In comparison, the one-loop Feynman diagrams predict the amplitude to be $\sim g_s^4(kL)^{1/2}/(16\pi^2) \sim g_s^2 (kL)^{-1/2}$, where we have used our previous power counting rule $\alpha_s/(4\pi) \sim 1/(kL)$ in the discussion of the massive KK graviton. Obviously, the one-loop amplitude is suppressed by the additional factor of $(kL)^{-1/2}$ compared with the tree-level one. This result is starkly contrasted with that of the massive KK graviton $G$ considered in the present paper.

\appendix

\section{Additional Diagrams}\label{AddD}
Due to the gauge covariance of the right-handed top quark kinetic terms, we should have two extra diagrams (c) and (d) in Fig.~\ref{fig:FeynD} besides of the triangle diagrams considered previously in Ref.~\cite{Geng:2016xin}. In this subsection, we calculate their contributions to the effective $Ggg$ couplings according to the interacting Lagrangian in Eqs.~(\ref{LagB}) and (\ref{LagA}). The calculations of the $G\gamma\gamma$, $G\gamma Z$ and $GZZ$ couplings follow the same procedure.

Note that only the massive graviton couples to the right-handed top quark field strongly, so that the amplitude of the Feynman diagram (c) is
\begin{eqnarray}\label{M2a}
i{\cal M}_{(c)} &=& \frac{g^2 c_{tt}}{2\Lambda} {\rm tr}(T^a T^b) (C_{\mu\nu,\sigma\kappa} - \eta_{\mu\nu} \eta_{\sigma \kappa})  \nonumber \\
&\times& \frac{1}{(2\pi)^4} \int d^4l \left( \frac{ {\rm Tr}[\gamma^\kappa P_R (\slashed{l} + m_t)\gamma_\rho (\slashed{l}+ \slashed{k}_1 +m_t)]}{(l^2-m_t^2)[(l+k_1)^2 -m_t^2]} \right) 
\nonumber \\
 &\times& \epsilon^{s\,\mu\nu}(P) \epsilon^{\rho\,*}(k_1) \epsilon^{\sigma\,*}(k_2) \nonumber\\
&=& \epsilon^{s\,\mu\nu}(P) \epsilon^{\rho\,*}(k_1) \epsilon^{\sigma\,*}(k_2) \frac{i g^2 c_{tt} \delta^{ab}}{4\Lambda} (C_{\mu\nu,\sigma\kappa} - \eta_{\mu\nu} \eta_{\sigma\kappa}) \frac{1}{16\pi^2} \int^1_0 dx   \nonumber\\
&&  \left\{\left[4 x(1-x) (\eta^\kappa_\rho k_1^2 - k_1^\kappa k_{1\rho}) \right] \left(\frac{2}{\epsilon} - \gamma_\omega + \ln \frac{4\pi \mu^2}{ m_t^2 - x(1-x)k_1^2}\right)\right\}\,,
\end{eqnarray}
where ${\rm tr}(T^a T^b) = \delta^{ab}/2$, the tensor $C_{\mu\nu,\sigma\kappa}$ is defined as in the appendix of Ref.~\cite{Han:1998sg},
and $k_1$ denotes the momentum of one external gluon. 
Consequently, if we use the on-shell conditions $k_1^2 = 0$ and $k_1^\rho \epsilon^*_\rho(k_1) = 0$,  the above amplitude vanishes. The same argument applies to the amplitude (d) in Fig.~\ref{fig:FeynD}, except for the exchange $k_1 \to k_2$. Note  that the amplitude obtained in Eq.~(\ref{M2a}) satisfies the Ward Identities for the external gluon $k_1$, so that this result is quite general in the view of the gauge invariance of QCD. Therefore, these diagrams do not contribute.

\paragraph*{Acknowledgements}
We thank Professor~Kentarou Mawatari and Dr.~Chen Zhang for their valuable comments.
This work was supported in part by National Center for Theoretical Sciences,
MoST (MoST-104-2112-M-007-003-MY3 and MoST-107-2119-M-007-013-MY3), NSFC (11547008) and
   the National Science Centre (Poland) research project (DEC-2014/15/B/ST2/00108).

\end{document}